\documentclass[amssymb,aps,a4]{revtex4}
\usepackage{graphicx,color,comment,amsmath,mathtools}
\usepackage{url}
\usepackage{hyperref}
\usepackage{amsfonts}
\begin{document}

\title{Self-organized manifold learning and heuristic charting via adaptive metrics}

\author{D. Horvath}
\affiliation{Centre of Interdisciplinary Biosciences, Faculty of Science, Jesenna 5,
P.~J.~Safarik University, 04154 Kosice, Slovak Republic}

\author{J. Uli\v{c}n\'y, B. Brutovsky}
\affiliation{Department of Biophysics, Faculty of Science, Jesenna 5,
P.~J.~Safarik University, 04154 Kosice, Slovak Republic}

\begin{abstract}
Classical metric and non-metric multidimensional scaling (MDS) variants are widely known manifold learning (ML) methods 
which enable construction of low dimensional representation (projections) of high dimensional data inputs. However, their use is crucially limited to the cases when data are inherently
reducible to low dimensionality. In general, drawbacks and limitations of these, as well as pure, MDS variants become more apparent when the exploration 
(learning) is exposed to the structured data of high intrinsic dimension. As we demonstrate on artificial and real-world datasets, 
the over-determination problem can be solved by means of the hybrid and multi-component discrete-continuous multi-modal optimization heuristics. 
Its remarkable feature is, that projections onto 2D are constructed simultaneously with the data categorization (classification) compensating in part for the 
loss of original input information. We observed, that the optimization module integrated with ML modeling, metric learning and categorization 
leads to a nontrivial mechanism resulting in generation of patterns of categorical variables which can be interpreted as a heuristic charting. 
The method provides visual information in the form of non-convex clusters or separated regions. Furthermore, the ability 
to categorize the surfaces into back and front parts of the analyzed 3D data objects have been attained 
through self-organized structuring without supervising. 
\end{abstract}

\maketitle

Manifold learning (ML)~\cite{Ma2011} is a technical term for a group of techniques
developed to reduce dimensionality of high-dimensional data, facilitating
their eventual visualization, evaluating and understanding at intuitive level.
The ML approach usually serves as a preprocessing step in familiarizing with 
data and for the formulation of hypotheses leading to further data analysis.
It has found many important applications in biology, robotics or visual data mining.
In recent years the ML techniques have also found applications in the
physics of deterministic chaos~\cite{Suetani2012}, as well as techniques for extracting 
structural information from X-ray diffraction snapshots~\cite{Schwander2012}. 

Projection from a higher to lower dimension is not straightforward and must meet
the ultimate requirement of capturing the essence or patterns of information
content of high dimensional datasets. Hence, the ML methods differ markedly in
the kind of the original information they are required to preserve during
the transformation. From this point of view, the basic classification of 
ML methods into local and global should be mentioned. 

The representative of the global methods are well known {\em principal component
analysis}~(PCA)~\cite{Jolliffe1989,Trendafilov2006} or 
classical MDS \cite{Kruskal1964a,Kruskal1964b} {\em non-linear} {\em
multidimensional} {\em scaling}~(MDS)~\cite{Cox1994,Cox2003,Carroll2005}.
The latter approach, MDS, is studied and extended in the present work.
Both the MDS variants, non-metric as well as metric, are formulated
as stress minimization problems where the stress is defined in the terms
of differences between pairwise dissimilarities of data points and distances of 
assigned projected coordinates~\cite{Chen2013}. Many methods have been developed 
that incorporate the conservation of the quantities such as distances and generalized distances.
In situations where the topological concepts offer more feasible
projections, pure Euclidean description is abandoned in favor of more flexible
geometries. For example, the method Isomap~\cite{Tenenbaum2000,Zha2007} 
uses geodesic instead of Euclidean distances and can be thus preferentially 
applied to nonlinear manifolds. 

Elementary ML methods face serious difficulties when confronted with
noisy~\cite{Yina2008} or intrinsically irreducible high-dimensional datasets.
In such cases, projection composed of approximate local isometrics is
usually constructed. In many real-world applications, one often tackles
general manifolds, where the domain decomposition and segmentation problems
often occur. These are typical for the closed manifolds, intersecting
circles, sphere surfaces or non-orientable surfaces such as M\"obius strip
or Klein bagel, which are analyzed later in this paper. The ML segmentation tasks 
producing non-overlapping domain decomposition are also known 
as identifications of {\em charts} and {\em atlases}~\cite{Duc2013}. 
The decomposition is NP hard local categorical assignment, which 
occurs in the graph coloring or graph partitioning problems. 

Until now, many alternative interdisciplinary approaches based on different principles 
have been developed to deal with the ML data preprocessing. The famous 
{\em stress} {\em function} {\em concept} has been first introduced by Kruskal~\cite{Kruskal1964a,Kruskal1964b}. 
Consequently, the scientists ~\cite{Sammon1969} are turning their attention
to new ideas and methods derived from the connection of ML with mathematical and physical modeling or 
nature-inspired optimization sciences~\cite{Kramer2012}. 
As an example may serve physically inspired ML method~\cite{Gorban2009}. 
Another method called {\em elastic} {\em map} exploits the mechanical 
analogy with the system of the elastic springs. The concept called
{\em diffusion map}~\cite{Nadler2005,Nadler2006} introduces diffusion distance 
less susceptible to the noise compared to Euclidean measures.
The approach {\em relational} {\em perspective map}~\cite{Li2004} consists 
in using parallels between mutual relations of data items and the behavior
of the positions of the charged particles, which are repelling each other
but are simultaneously confined to toroidal surface. Several 
new approaches to perform ML~\cite{Xiao2011} including kernel
regression~\cite{Kramer2012} take inspiration from the evolutionary 
general-purpose heuristics and genetic algorithms. 
In~\cite{Chen2013} the class of the models based on the generalized B-C energy (stress) 
functions~\cite{Noack2007,Noack2009} which has its origins 
in the optimization approach of Sammon~\cite{Sammon1969}. 

Rapidly increasing computer power allows to tackle ML problems in previously
unexpected ways. Further progress in the ML and MDS techniques can be made
by applying simple, but computationally demanding {\em optimization-based} {\em approaches}.
In this respect, we propose the variant of data adaptive metrics aimed to
provide combined description in the terms of continuous/quantitative and
categorical variables. Simultaneous use of discrete (categorical) and continuous variables 
needs heuristics-based optimization to avoid getting stuck in the local optima. 
As demonstrated in the below presented simulation examples, the advanced hybrid 
construction enables to adapt the distance metrics simultaneously 
with the categorical classification and adjustment of the continuous projected coordinates.

The ML process may be also viewed as a kind of stochastic optimization 
which is inspired by the imitation of natural systems. Designers of optimization techniques usually 
imply biologically-inspired concepts~\cite{Alatas2010} to discover suitable rules~\cite{Binitha2012}.
On the other hand, many heuristics-based optimization methods benefit from
the analogies between the optimization dynamics and physical processes as well~\cite{Biswas2013}.

In the paper we present computational results of heuristic simulation
MDS technique. We assume, that the combination of a few existing optimization
approaches may provide better results than the only method.
The stochastic optimization method we applied combines the advantages
of {\em grid} {\rm search}~(GS)~\cite{Horst1995}, {\em extremal} {\rm optimization}~(EO)
\cite{Boettcher1999}, and {\em hysteretic optimization}~(HO)~\cite{Zarand2002}. 
As demonstrated below, dynamical behavior of the above combination produces very interesting behavior.
In the next we give a brief description of the respective methods and their benefits. 

GS is the standard way of performing exhaustive optimization in the hyperparameter space. 
But this strategy does not scale well for large problems. GS is an efficient in a one-dimensional 
or two-dimensional domains since the problems at higher dimensions occur due to "the curse of dimensionality". 

The problem of getting stuck in local optimum is partially solved by incorporating
EO method, inspired by the stylized model of coevolutionary process proposed by Bak
and Sneppen~\cite{Bak1993}. Since then, many authors have extended the method
(see e.~g.~\cite{Boettcher2000}) and applied it in many contexts~\cite{Zhou2005}.
The essence of the method consists in the identification of the low-quality components
and their subsequent elimination. The method exploits highly nonlinear mechanism
of the large fluctuations - avalanches, known to be efficient in the exploration of many local optima 
and overcoming of the barriers in the search space~\cite{Boettcher2000P}. 
The main difficulty with the EO applications is, that its implementation necessitates specific definition
of the local fitness (scoring, objective) function.

HO method~\cite{Pal2006} is inspired by the mechanism of the global reordering
during the demagnetization of magnetic samples due to damped alternating magnetic field. 
Nevertheless, the method can be formulated more abstractly and adapted to non-magnetic 
problems as well. The HO method provided successful outcomes in the case 
of the benchmark {\em traveling} {\em salesman} {\em problem}~\cite{Pal2003}.
We justify the below presented ML application of the HO type technique
by the fact, that suggested type of distance metrics involves global parameter
with system-wide impact which can be roughly regarded as analogous
to the intensity of external magnetic field.

The paper is organized as follows. In the section~\ref{sec:Mod} we describe 
the formulation of MDS with the use of the adaptive metrics. In section \ref{sec:optim}
we discuss the optimization strategies appropriate for given purpose. 
The datasets and corresponding numerical results illustrating 
our approach are described in sec.\ref{sec:numerical}. 
Finally, the conclusions are presented. 

\section{MDS with adaptive metrics}\label{sec:Mod}

Below we analyze N data items, each having $D$ ($D>2$) components (column features, or classes)  
\begin{equation}
\{ \, {\bf X}_i \in \mathbb{R}^{D} \,, \,\, i =  1, 2, \ldots, N \,\}\,,    
\label{eq:Xis}
\end{equation}
where ${\bf X}_i = [\,X_{i,1},$ $X_{i,2},$ $\ldots, $ $ \, X_{i,D} ]$.
Regarding MDS technique, essential information is comprised in $N\times N$
elements of dissimilarity matrix $d^{(D)}_{i,j}$. In here 
presented specific application, $ d^{(D)}_{i,j}$ obtains standard 
Euclidean form $ d^{{\rm E},(D)}_{i,j} = $ $\sqrt{(1/D)\sum_{z=1}^D (X_{i,z}-X_{j,z})^2}$,
but other choices are possible as well. 

The process of dimensional reduction onto dimension $P<D$ can be viewed
as ongoing iteration of the configuration tuples
$C(t) \in \mathbb{R}^{P} \times \Omega \times \mathbb{R}^N$
including $N$ data points 
\begin{eqnarray}
C(t)  &  \equiv & \left\{\, \, 
\left[{\bf x}_{1}(t), s_{1}(t) \right], 
\left[{\bf x}_{2}(t), s_{2}(t) \right], \ldots, 
\left[{\bf x}_{N}(t), s_{N}(t)\,\right] \,\right\}\,, 
\label{eq:virtpart1}
\\
{\bf x}_{i}(t) & \in  & \mathbb{R}^P\,, 
\nonumber 
\\
s_{i}(t) &  \in  &  \Omega  \equiv \{\, 0, 1,  \ldots, N_{\rm s} -1 \,\} 
\nonumber
\end{eqnarray}
where $i = 1, 2, \ldots, N$; the discrete time $t$ ranges from $0$ to $t_{\ast}$;
$C(t)$ consists of the system of $N$ vectors of $P$ real valued Cartesian
coordinates $x_{i,1}(t)$, $\ldots$, $x_{i, P}(t)$.
One of the cornerstones of the proposed approach is, that the uncertainty
and frustration which arose from a projection effect may be reduced by introducing categorical 
variables $s_i(t)$, $N_{\rm s}$ being the number of their possible values. 

At the heart of the MDS ML approach stands the requirement of approximate fulfillment
of $N(N-1)/2$ conditions after the stop time $t_{\ast}$
\begin{equation}
d_{i,j}^{(P)}(t^{\ast})\equiv d^{(P)} 
\left( 
{\bf x}_i ( t_{\ast}), 
s_i(t_{\ast}), {\bf x}_j(t_{\ast}), s_j(t_{\ast})\right)  
\simeq  d_{i,j}^{(D)}\,,
\label{eq:distcons}
\end{equation}
which approximate $d^{(P)}_{i,j} = d^{(D)}_{j,i}$.
Since the conditions of distance-preservation are too demanding to be achieved
in all the eventual applications, it is desirable to solve an approximation
problem by iterative optimization. The quality of the approximation 
may be assessed through the absolute error term 
\begin{eqnarray}
e_{i,j}(t) & \equiv & {\Bigg|}\, \frac{d_{i,j}^{(P)}(t) - d_{i,j}^{(D)}}{d_{i,j}^{(D)}}\,{\Bigg|}
\nonumber 
\label{eq:err}
\end{eqnarray}
but more appropriate case-dependent scoring variants and weighting schemes may be
devised for specific situations. The effort is to achieve trajectories revolving
around desired outcome 
$ e({\bf x}_i(t_{\ast}),$ $s_i(t_{\ast}),$ ${\bf x}_j(t_{\ast})),$ $s_j(t_{\ast}))$ $= 0$.

In analogy with the well known additive interaction effects we assume, that 
required properties of the $i$-th projected component may 
be attained by checking the values of the local 
potentials constructed as 
\begin{equation}
V_i (C) =\frac{1}{N-1} \sum_{j=1; \,j\neq i}^N e_{i j}\,. 
\label{eq:Vi}
\end{equation}
The overall views about the system performance and convergence can 
be obtained by minimizing the total potential
\begin{equation} 
V_{\rm tot}(C)= \frac{1}{N} \sum_{i=1}^N V_i(C)\,. 
\label{eq:Vtot}
\end{equation} 
Note, that the term stress function is more commonly used within the ML MDS
context~\cite{Sammon1969,Chen2013}. When seen from the point of view of Bak-Sneppen model~\cite{Bak1993}, 
the value $V_i$ plays role of the fitness. Being inspired by Monte Carlo simulations of the spin systems, 
the overall categorization dynamics was characterized by calculating the instant "magnetization" 
\begin{equation}
{\rm Mag}(t) = \frac{1}{N} \sum_{i=1}^N s_{i}(t)\,.
\end{equation}
In the presented version 
of MDS algorithm we propose parametric distance measure 
\begin{eqnarray}
d^{(P)}({\bf x}_i, s_i, {\bf x}_j, s_j) = 
\left(\,1+ H | s_i - s_j |\,\right) d^{{\rm E},(P)} ({\bf x}_i, {\bf x}_j)\,.  
\label{eq:Hell}
\end{eqnarray}
Instead of relying on pure Euclidean distance $d_{i,j}^{{\rm E},(P)}$, we use modification 
with the multiplicative factor $1 +  H | s_i - s_j |$ that is supposed to improve the 
matching according Eq.(\ref{eq:distcons}). Here, $ H\in \langle H_{\rm D},
H_{\rm U}\rangle \subset \mathbb{R}$ is the real valued parameter. Its global
system impact motivates the use of the HO optimization. The dependence 
on $| s_i - s_j |$ represents the interaction due to differences in categories. 
When the different data items are differently categorized ($s_{i\neq j} \neq s_j$),   
the Euclidean distance $d_{i,j}^{{\rm E},(P)}$ changes in 
the positive or negative sense according to the sign of selected $H$
parameter.  The matching of the categories ($s_i = s_j$) simply yields 
basic choice $d_{i,j}^{(P)}= d_{i,j}^{{\rm E},(P)}$.
Since not only $ \{\,{\bf x}_i \}_{i=1}^N$, but $\{ s_i \}_{i=1}^N$ and
$H$ are unknown as well, the approach constitutes complex inverse problem
which requires simultaneous tuning of distance, category and metrics.
This optimization problem is solved by combining beneficial features
of the three mentioned optimization methods: GS, HO and EO.
As demonstrated in our numerical experiments, the iterative procedure
incorporating them can exhibit very complex dynamics and behaviors.
The optimization methods are considered to have access to different
subsystems: (i)~GS method is applied to optimize $\{  {\bf x}_i\}_{i=1}^N$
and $\{ s_i \}_{i=1}^N$; (ii)~the parameter $H$ is optimized by the {\em self-organizing}
{\em dynamics} based on the modified HO method; (iii)~EO applied to vary 
$\{ s_i \}_{i=1}^N$ to disentangle partially improperly justed categories.
The self-organization arises through decentralized interactions without primary knowledge of the way 
how to redistribute the information from higher dimension among 
the discrete ($\{s_i\}_{i=1}^N$), continuous ($\{x_i\}_{i=1}^N$, $H$) degrees of freedom. 
This form of learning is often referred to as unsupervised learning or classification.

As complexity of the embedded data increases, it is unfeasible to design metrics from scratch. 
We believe that the appropriate MDS design starts with the
definition of adaptive and local (determined by $s_i$) geometry,
such as that defined by Eq.(\ref{eq:Hell}). Our attempt was, from methodological viewpoint, 
partly inspired by theoretical framework of general relativity 
and geometrodynamics, where the fundamental postulate is made that geometry is determined by the mass-energy 
distribution analogous to distribution and structure of high dimensional dataset as counterpart. 

In the field of MDS research, we would like to mention at least two 
approaches to "distance metric learning" 
or distance adaption particularly close in motivation to our
approach. In~\cite{Kari2013} the distance is replaced by the linear function with the 
parameters determined by the regression. In the second approach~\cite{Cansizoglu2014} 
the monotonic nondecreasing function of the distance has been introduced in
order to make the differences of distances less significant. In addition, our  approach 
can be considered as being in line with the class of the adaptive ML approaches discussed in~\cite{Yin2011}. 
The results from implementation of {\em metric learning} approaches should be mentioned as well~\cite{Zhang2014}. 

\section{Combination of particular optimization strategies}\label{sec:optim} 

Below we present the heuristic optimization algorithm tailored to solve the MDS
problem with adaptive metrics. The algorithm stops at time $t^{\ast}$ and
consists of the following subsequent steps (enumerated by $t$)

\vspace*{4mm}

\noindent \underline{{\it Step 1}}: {\bf GS optimization in polar coordinates}  

To refine the optimum estimation locally, we use polar grid mesh around randomly
localized ${\bf x}_{i_{\rm rand}}$ with $i_{\rm rand}$ drawn uniformly randomly 
from the set $\{\, 1,  2,  \ldots,  N\,\}$.
The mesh is created with the radial step resolution $\Delta r$ and angular step
resolution $ 2\pi /N_{\rm n}$. The mesh parameters $\Delta r$ are drawn uniformly 
randomly from the respective interval $ \langle  \Delta r_{\rm D},
\Delta r_{\rm U} \rangle$. 
Then in the special case considered here $P = 2$ 
the algorithm generates 
the mesh of $N_{\rm n}^2 N_{\rm s}$ nearest-neighbor polar grid points 
\begin{eqnarray} 
x_{i_{\rm rand},\,1}^{\rm cand}(l_{\rm r}, l_{\phi}) &=& 
x_{i_{\rm rand},\,1} +  \frac{\Delta r\, l_{\rm r}}{N_{\rm n}} \cos\left(\frac{2 \pi l_{\phi}}{N_{\rm n}} \right)\,,
\\
x_{i_{\rm rand},\,2}^{\rm cand} (l_{\rm r}, l_{\phi}) &=&  x_{i_{\rm rand},\,2} + \frac{\Delta
r\, l_{\rm r}}{N_{\rm n}} 
\sin \left(\frac{2 \pi l_{\phi}}{N_{\rm n}} \right) \,,
\nonumber 
\\
s_{i_{\rm rand}}^{\rm cand} (l_{\rm s})  &= & l_{\rm s}\,. 
\nonumber 
\end{eqnarray}
Within the standard logic of GS approach, the projections $\{\, x_{i_{\rm rand}, z}^{\rm cand} ( l_{\rm r}, l_{\phi}),\,\, 
z = 1, 2\}$, denoted by the superscript 'cand', represent 
candidate solutions of the respective optimization problem. 
Then, the candidate projections are enumerated by the triplets
\begin{equation}
\{ \left( l_{\rm r},\, l_{\phi}, \, l_{\rm s}\right);\, l_{\rm r}, 
l_{\phi} =  1, 2, \ldots,  N_{\rm n}\,;\,l_{\rm s} = 0, 1, \ldots, N_{\rm s}-1\,\}\,.
\end{equation}
In the case of feasibly high $ N_{\rm s}$ and $N_{\rm n}$, one can explore all
the possible categories of $s_{i_{\rm rand}}$ combinatorially.
Obviously, the calculation of $V^{\rm cand}_{i_{\rm ran}}=V_{i_{\rm ran}}$ using 
Eq.(\ref{eq:Vi}) must be preceded by reevaluation of the 
distances from ${\bf x}^{\rm cand}_{i_{\rm ran}}$,  $s^{\rm cand}_{i_{\rm rand}}$
to all the other points. 
Let the coordinates $x_{i_{\rm rand},1}^{\rm cand}(l_{{\rm r},{\min}},
l_{{\phi},{\min}})$, $x_{i_{\rm rand},\,2}^{\rm cand}(l_{{\rm r}, {\min}},
l_{{\phi},{\min}})$, $s_{i_{\rm rand}}^{\rm cand}(l_{{\rm s}, {\min}})$ 
correspond to the lowest local value of $V^{\rm cand}_{i_{\rm rand}}$.
As this value is calculated using Eq.(\ref{eq:Vi}), its calculation must
include changes in $e_{i_{\rm rand},j}$ and $ d_{i_{\rm rand},j}^{(\rm P)}$
which are needed to update the ${\bf x}_{i_{\rm rand}}$ and $ s_{i_{\rm rand}}$
values used in further optimization iterations. 

\noindent \underline{{\it Step 2}}: {\bf EO in the space of categorical variables}  

\vspace*{1mm}

The optimization step is accepted with the decaying probability $ \exp(-t/t_{\rm dec})$
suggested to decay in time with the characteristic time constant $t_{\rm dec}$. 
The strategy is similar to simulated annealing approach.
At each algorithmic step, the instant worst part  $i_{\rm max}  
\in  \{\, 1, 2, \ldots,\, N\}$ of the system defined by the 
respective
maximum  $V_{i_{\rm max}} = \max_{i \in \{1, 2, \ldots, N\} } V_i$ is localized.
Then, the categorical variable $s_{i_{\rm max}}$ is replaced by the value
of $s_i$ drawn randomly from the set $\{\, 0,   1, \ldots, N_{\rm s} - 1\,\}$. 

\vspace*{4mm}

\noindent \underline{{\it Step 3}}: {\bf HO - hysteresis along the variable $ H $}

\vspace*{1mm}

Let's denote the best estimate of the optimum of the stress that algorithm
attained by the time $(t-1)$ as $V_{\rm tot, best}(t-1)$ and the total
instant tension (potential) attained in time $t$ as $V_{\rm tot}(t)$, 
both calculated using Eq.~(\ref{eq:Vtot}) at the respective times. 
Let $V_{\rm tot,best}(t-1)$ corresponds to $H_{\rm best}(t-1)$ 
estimate of $H(t^{\ast})$. Then, if $V_{\rm tot, best} (t-1) > V_{\rm tot}(t)$, the
algorithm updates the $V_{\rm tot,best}(t)$ and $H_{\rm best}(t)$ as 
\begin{eqnarray}
H_{\rm best}(t)  \leftarrow  H(t)\,,\qquad V_{\rm tot, best}(t)  \leftarrow  V_{\rm tot}(t)\,.
\end{eqnarray}
Otherwise, previously obtained values are used to update the $V_{\rm tot, best}$ and $H_{\rm best}$, 
respectively
\begin{eqnarray}
H_{\rm best}(t) & \leftarrow &  H_{\rm best}(t-1)\,,  
\qquad  
V_{\rm tot,best}(t)  \leftarrow V_{\rm tot,best}(t-1)\,.
\end{eqnarray} 
The HO dynamics is driven by the periodic exogenous signal 
\begin{equation}
H_{\rm per}(t) = H_{\rm D} +  \frac{H_{\rm U} -  H_{\rm D}}{2} 
\left[\, 
1 + \cos \left( \frac{2 \pi t}{t_{\rm per}}\right)\right]
\end{equation}
with the properly chosen period $t_{\rm per}$. The signal $H_{\rm per}(t)$ represents oscillations bounded 
by the $H_{\rm D}$, $H_{\rm U}$ constants. The algorithm applies the strategy of the sequential linear mixing
of the best estimate of the optimum with the decaying oscillations. 
The mixing is characterized by the coefficient $\exp \left(-t /  t_{\rm dec} \right)$. 
The mixing process is incorporated into the non-autonomous 
recurrent dynamic rule in the form  
\begin{equation}
H(t+1) = \underbrace{ H_{\rm per}(t)  + 
\left(\,  H_{\rm best}(t)  -  H_{\rm per}(t)\,\right) 
\underbrace{\left[ 1 - \exp \left( - \frac{t}{t_{\rm dec}}\,
\right)\right]}_{\mbox{\small tends to $1$ as
$t\rightarrow\infty$}}}_{\mbox{\small tends to $H_{\rm best}(t)$ as $t\rightarrow \infty$}}\,. 
\label{eq:Hlearn}
\end{equation}
We remark, that an analogous learning strategy has 
been followed earlier~\cite{Horvath2004}
to solve the problem of Monte Carlo localization of the critical point under
the noisy conditions.  By other words, the formula is designed to reduce coupling between
$H(t+1)$ and $H_{\rm per}(t)$ under increasing influence of $H_{\rm best}(t)$. 
It is straightforward to expect, that the ability to localize
(hopefully global) optimum $H(t^{\ast}+1) \sim H_{\rm best}(t^{\ast})$
needs approximate fulfilment 
of the condition $t^{\ast} \gg   t_{\rm dec} \gg t_{\rm per}$. 

\section{Numerical experiment}\label{sec:numerical}

To illustrate here proposed algorithm, we apply it to several datasets drawn from analytically specified low dimensional manifolds embedded in
an ambient space. 

Firstly, we specify the list of parameters, which are common for all the system optimizations. The MDS with the projection onto $P=2$ dimensions
has been performed for $t^{\ast} \sim 3.0 \times 10^6$ iteration steps, but smaller number of iterations is sufficient for achieving comparable quality
of the 2D projections.  In the most cases, we consider systems including three data categories $N_{\rm s}=3$, 
thus $s_i  \in \Omega= \{ 0, 1, 2 \}$ (in the specific cases we used $N_{\rm s}=4$, $5$, $6$; the detailed specification is given 
in the corresponding figure captions). The optimization has been done for the search parameters 
$N_{\rm n}=10 $ and $H_{\rm D} = -1.5 $, $H_{\rm U} = 1.5 $ (the bounds $H_{\rm D}=-3.5$, 
$H_{\rm U} = 3.5$ were also used to verify stability of obtained results). To perform the GS strategy, the mesh 
size was left to fluctuate  within the bounds $\Delta r_{\rm D} = 0.001$, $\Delta r_{\rm U} = 0.3$.
The dynamics of $H(t)$ was determined by the exogenous signal characterized by the parameters $ t_{\rm dec} = 3 \times 10^5$ and $t_{\rm per} = 24 \times 10^3$.
The optimization has been initialized from $ x_{i,0} = X_{i,0}$,  $x_{i,1} = X_{i,1}$ with the small additive noise, but the numerical experiments revealed
that the initial conditions have only negligible influence on 2D projections. The tendencies of the heuristics and convergence towards 
the optimum has been controlled by monitoring of $V_{\rm tot}(t)$.  Any parametric approach requires parameter estimation. Despite many parameters to tune, we 
observed surprising robustness of the presented method in most cases as well as in different situations. The experience has shown 
us that what we need to focus on is the choice of parameters $t_{\rm dec}$, $t_{\rm  per}$, $H_{\rm D}$ and $H_{\rm U}$
which seems to us play a key role in the determination of the minima.

The constructiveness of the algorithm is shown on the example of dataset. 
First, we considered system of $N=300$ data items embedded into $D=6$ space. 
The data were drawn from the parameterization 
\begin{eqnarray}
X_{i, 1}   & = & \left[ 1 +  \frac{1}{2} \cos\left(\frac{16 \pi i}{N}\right) \right] \cos\left(\frac{2
\pi i}{N} \right)\,, 
\label{eq:torspiral}
\\
X_{i, 2}   & = &  \left[ 1 + \frac{1}{2} \cos\left(\frac{16 \pi i}{N}\right) \right]   
\sin\left(\frac{2 \pi i}{N}\right)\,,
\nonumber
\\
X_{i,3}  & = & \frac{1}{2} \sin\left(\frac{16 \pi i}{N}\right)\,, 
\nonumber 
\\
X_{i,4} & = &   a_{\rm m}\, \delta_{0, i \,{\rm mod} 3}\,, 
\quad 
X_{i,5} =     a_{\rm m}\, \delta_{1, i \,{\rm mod} 3}\,,
\quad 
X_{i,6} =     a_{\rm m}\, \delta_{2, i \,{\rm mod} 3}\,.  
\nonumber 
\end{eqnarray}
The above dataset is constructed as a combination of the {\em toroidal
spiral} (coordinates $X_{i, 1}, X_{i, 2}, X_{i, 3}$) modified by geometric effects added by Kronecker
delta and modulo functions $\delta_{j, i {\rm mod} 3}$ calculated for the $j = 0, 1, 2$ components. 
The variable $a_{\rm m}$ is used to study different 
optimization conditions. 

\begin{figure}
\centering
\includegraphics[width=1\linewidth]{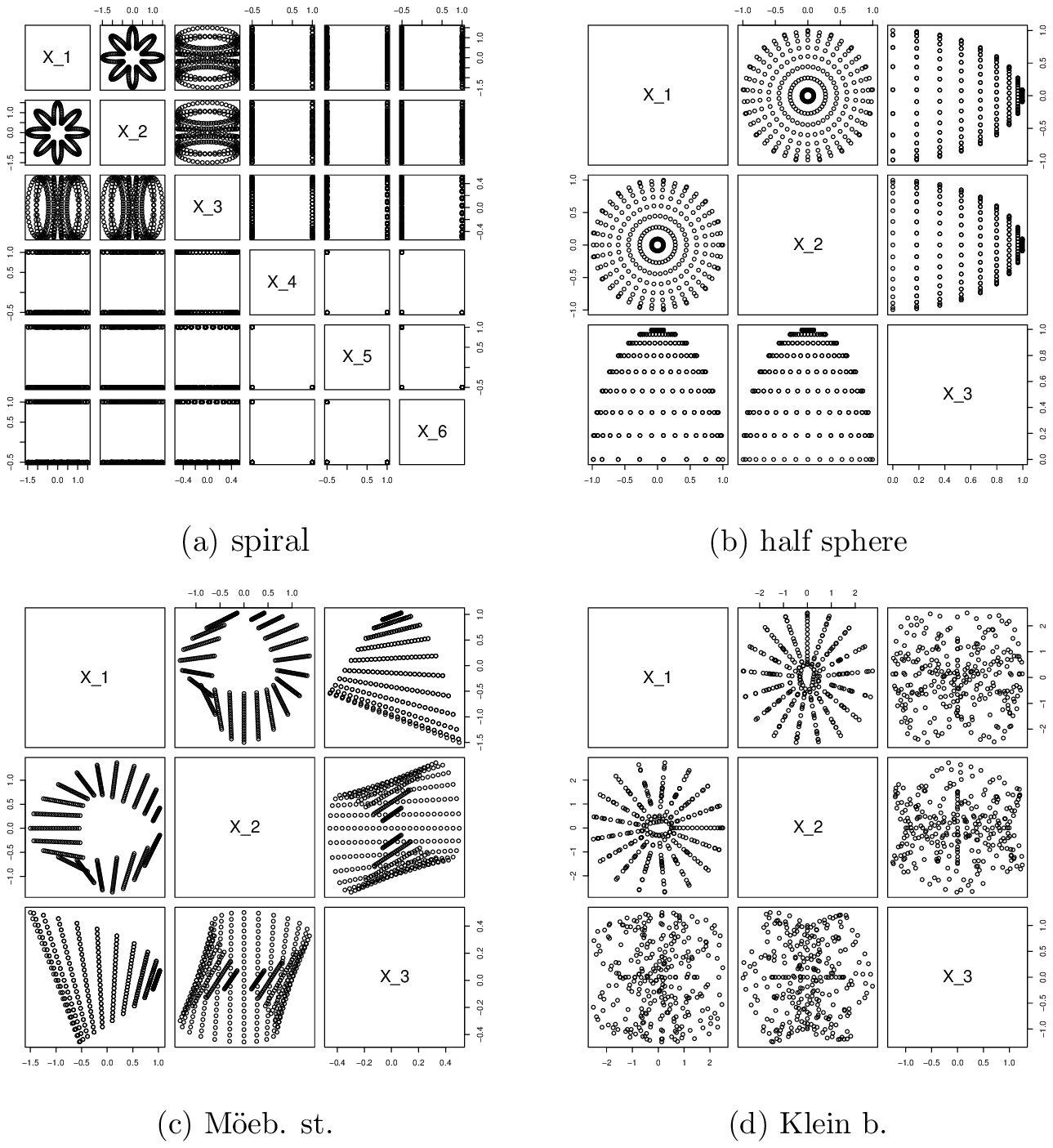}      
\caption{The multiple scatterplot attempt to represent the relationships among variables of
four selected artificial datasets. The data for 6D modulated toroidal spiral are drawn 
from Eq.(\ref{eq:torspiral}) for the parameter $a_{\rm m} = 1.2$.}
\label{Fig1}
\end{figure}

\begin{figure}
\centering
\includegraphics[width=1\linewidth]{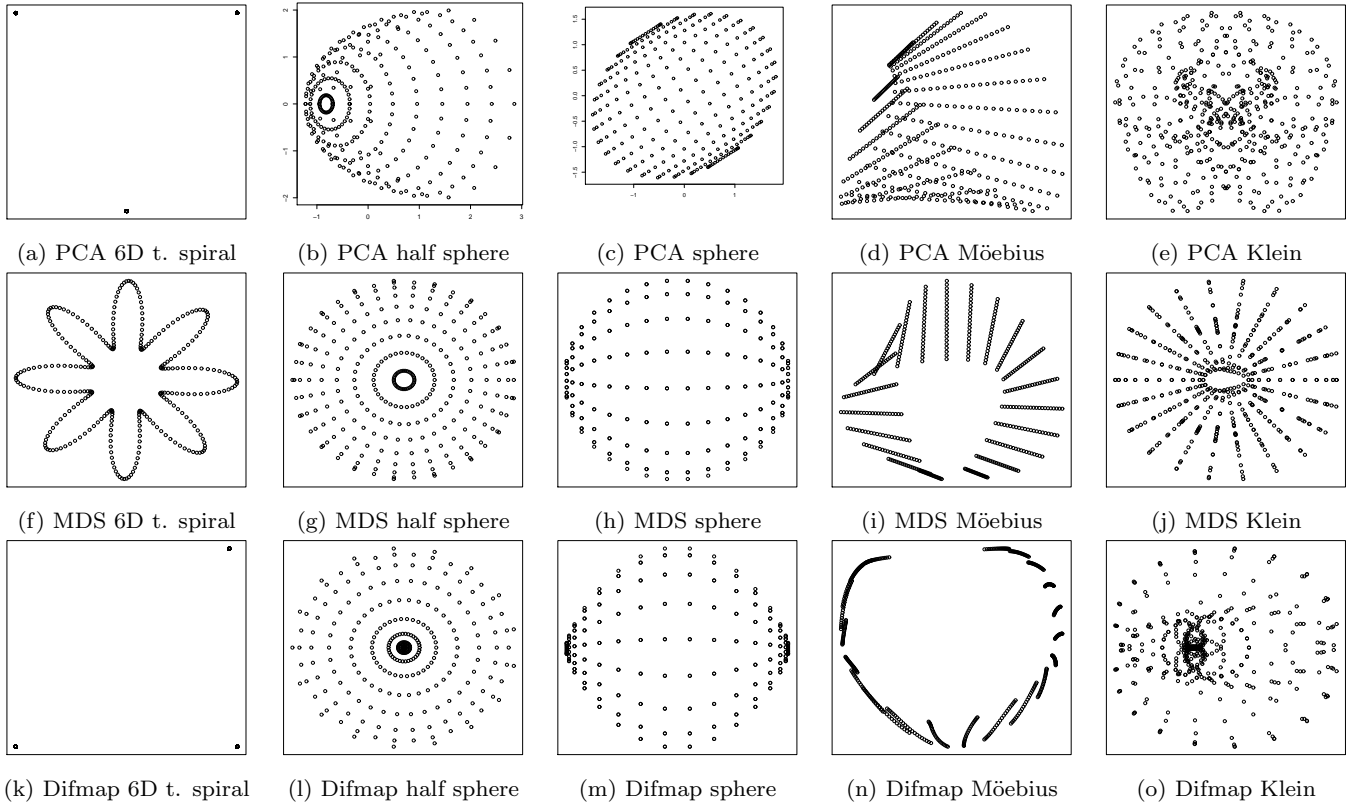}      
\caption{The comparison of the PCA, MDS and diffusion map is done for several 
generated artificial datasets. It should be noted that diffusion maps 
rely on the choice of the tuning parameter $\epsilon$ that dictates 
the threshold for similarity. We present the results for $\epsilon=0.1$. 
Surprisingly, many of the phase portrait patterns are reminiscent of those 
found in Fig.\ref{Fig1}. The most remarkable differences between outputs of the mentioned methods have
been detected for 6D toroidal spiral ($a_{\rm m}=1.2$) and Klein bagel.} 
\label{Fig2}
\end{figure}

In Appendix I we present formulas for generating supplementary artificial datasets 
which outline consequences of the proposed method. 
In order to gain preliminary understanding of datasets, the relations between
data pairs of Cartesian coordinates is plotted in Fig.\ref{Fig1}.
In addition, to evaluate and facilitate the understanding of our method we compared
results for PCA, classical metric MDS and diffusion map. 
The results are shown in Fig.\ref{Fig2}. Let us to note that in the case of PCA and MDS we used {\rm princomp}() 
and {\rm cmdscale}() R's base functions from {\rm stats} package. For the implementation of diffusion map 
we used R function {\rm diffuse}() from {\rm diffusionMap}~\cite{RDifRich2014} package. 

\begin{figure}
\centering
\includegraphics[width=1\linewidth]{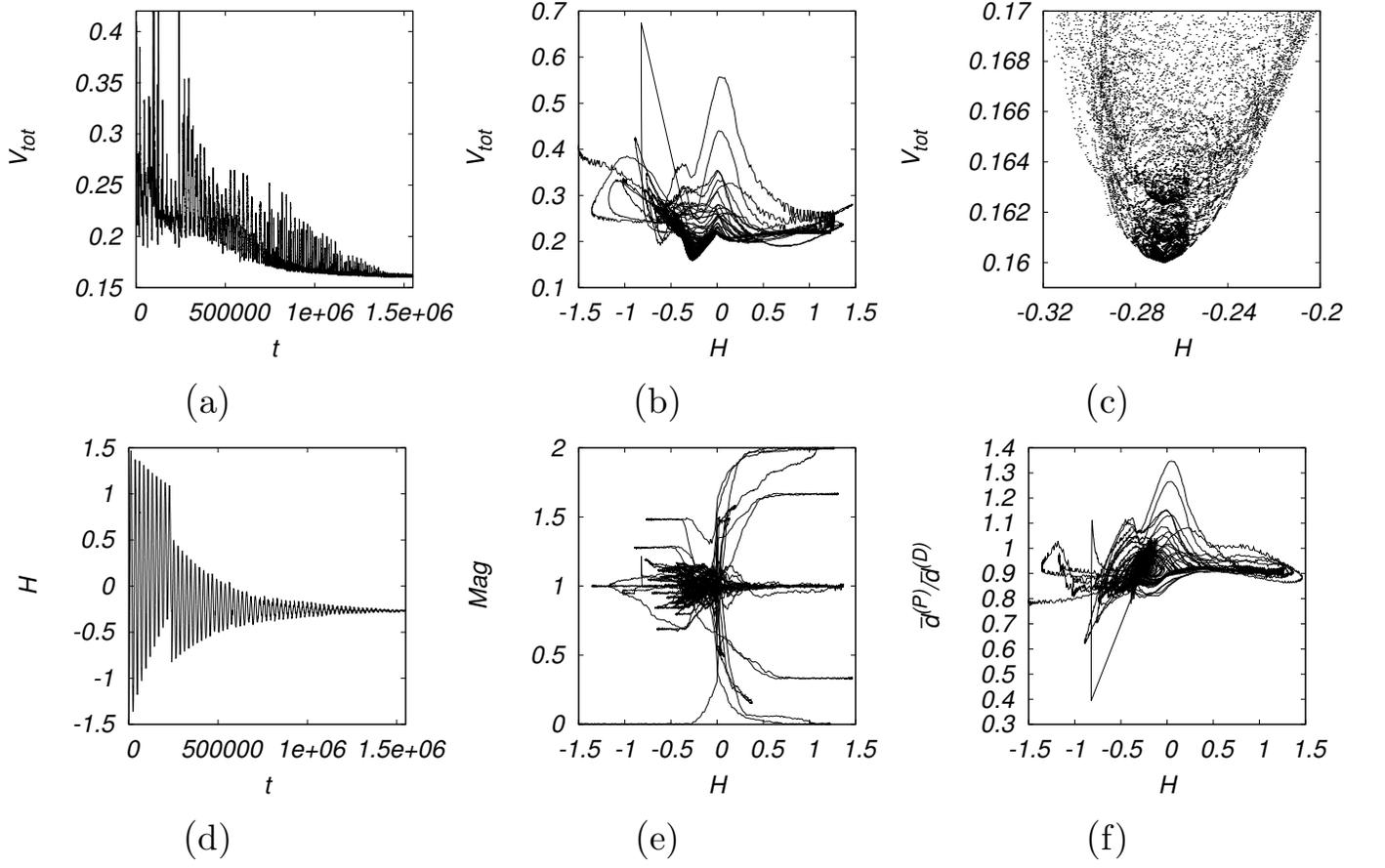}      
\caption{The MDS optimization represented by phase portraits in terms of collective variables. 
The formation of MDS projection of the original data sampled from 6D toroidal spiral 
defined by Eq.(\ref{eq:torspiral}); the artificial data calculated for $a_{\rm m}=1.5$. 
The dependences show remarkable effect of the exogenous oscillations. 
The figure parts inform how the particular variables vary during optimization process. 
The part~(a) demonstrates global decrease of $V_{\rm tot}(t)$.  The visible are random excitations 
(avalanches) which stem from the application of EO. The figure parts~(b), (c) depict the parametric 
plot of the total stress $V_{\rm tot}(t)$. They admit to estimate the optimal 
$H(t^{\ast})$ that corresponds to $V_{\rm tot}(t^{\ast}) \sim -0.27$.
The part (d)~depicts the illustrative break in the decaying exponential regime of $H(t)$ caused by the 
self-organized HO. The part (e)~shows that at the minimum the instant
$Mag(t)$ saturates around 1; (f)~The alternative confirmation 
of the convergence of the optimization: the optimized ratio of the mean distances 
of the original and projected coordinates 
tends to the value slightly less than 1.}
\label{Fig3}
\end{figure}

\begin{figure}
\centering
\includegraphics[width=1\linewidth]{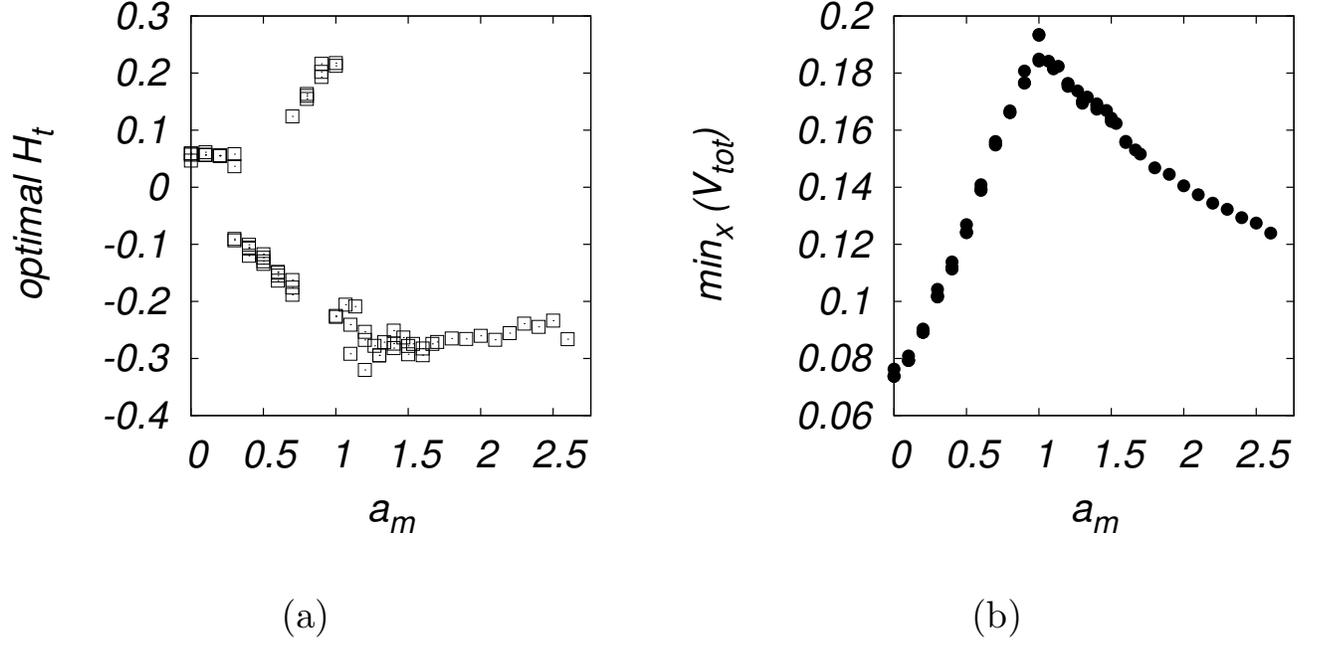}      
\caption{The systematics of the of optimal characteristics of the projections of 6D toroidal spiral. 
The results obtained for different $a_{\rm m}$ (see Eq.(\ref{eq:torspiral})). 
We see that data may also exhibit critical-like properties of nonlinear changes. 
The part~(a) shows several breaks in the $a_{\rm m}$ dependence of the optimal 
$H(t^{\ast})$. Two branches of the competing solutions differ in the sign of $H(t^{\ast})$. The part~(b) shows qualitative 
change indicated by the cusp-peak of the optimal $V_{\rm tot}(t^{\ast})$ localized near to $a_{\rm m}\sim 1$. 
The peak appears to be related to the structural change, which is viewable 
in the corresponding configuration  ($a_{\rm m}\simeq 1$) in Fig.\ref{Fig5}.}
\label{Fig4}
\end{figure}

\begin{figure}
\centering
\includegraphics[width=1\linewidth]{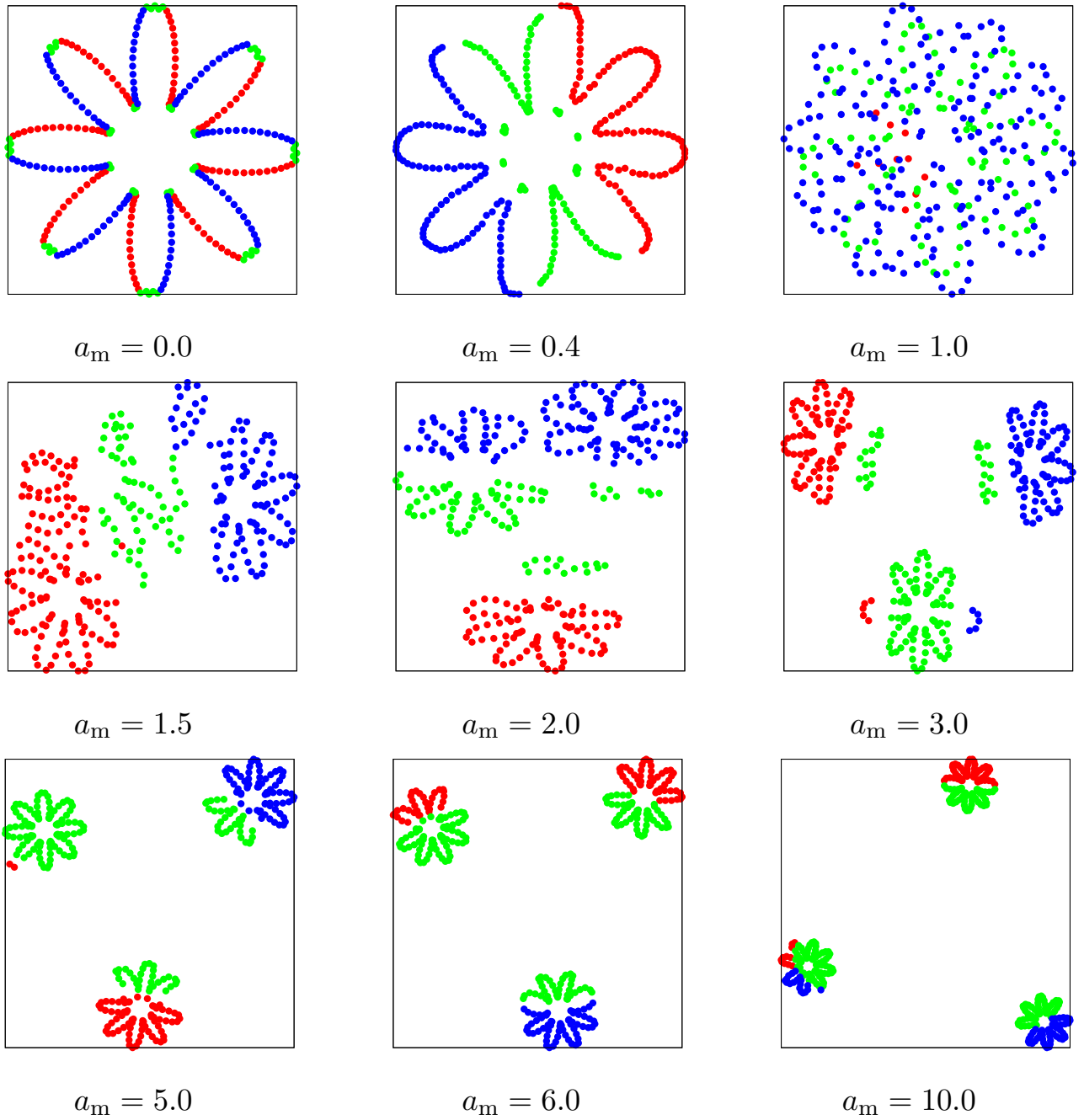}      
\caption{The nearly optimal configurations of the spiral projections 
obtained for different $a_{\rm m}$ values. The 
ternary like structure forms due to presence of the modulo function. 
Moreover, it is rather logical to observe that separation between 
three visible parts increases with $a_{\rm m}$. The results suggest different degree 
of the regularity and different quality of results as well for different $a_{\rm m}$.} 
\label{Fig5}
\end{figure}

Let us focus on the problem of spiral studied for different $a_{\rm m}$. The calculations 
(see the optimization results in Figs.~\ref{Fig3} and \ref{Fig4} and the corresponding
configurations in Fig.~\ref{Fig5}) revealed, that large $a_{\rm m}$ ($a_{\rm m} \gg 1$) 
enhances the segregation process due to higher impact of the modular data structure and smaller 
influence of the harmonic functions forming the 6D spiral. One of the most interesting findings is, that qualitative 
differences and regimes (see for example $a_{\rm m} \sim 1$ cases presented in Fig.\ref{Fig4} and Fig.\ref{Fig5}) 
may make the MDS analysis of some specific datasets more difficult than others. 

\begin{figure}
\centering
\includegraphics[width=1\linewidth]{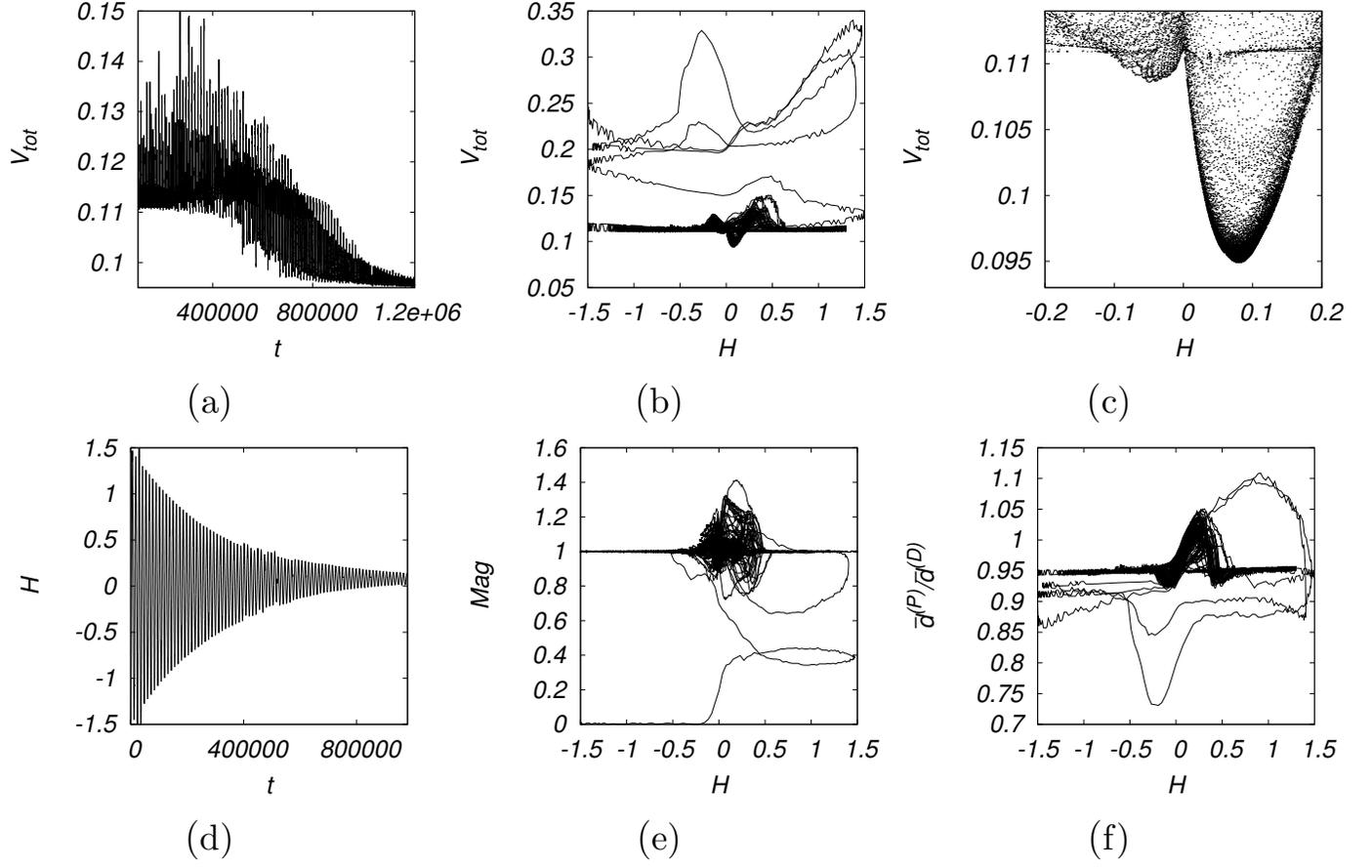}      
\caption{The monitoring of the complex optimization dynamics, which yields 
2d projection of the Klein bagel. The results highlight the inherent complexity of the
optimization dynamics. The stochastic attempts of overcoming of local barriers are clearly visible. 
See Fig.\ref{Fig3} for more detailed comments of the partial phase portraits.}
\label{Fig6}
\end{figure}

\begin{figure}
\centering
\includegraphics[width=1\linewidth]{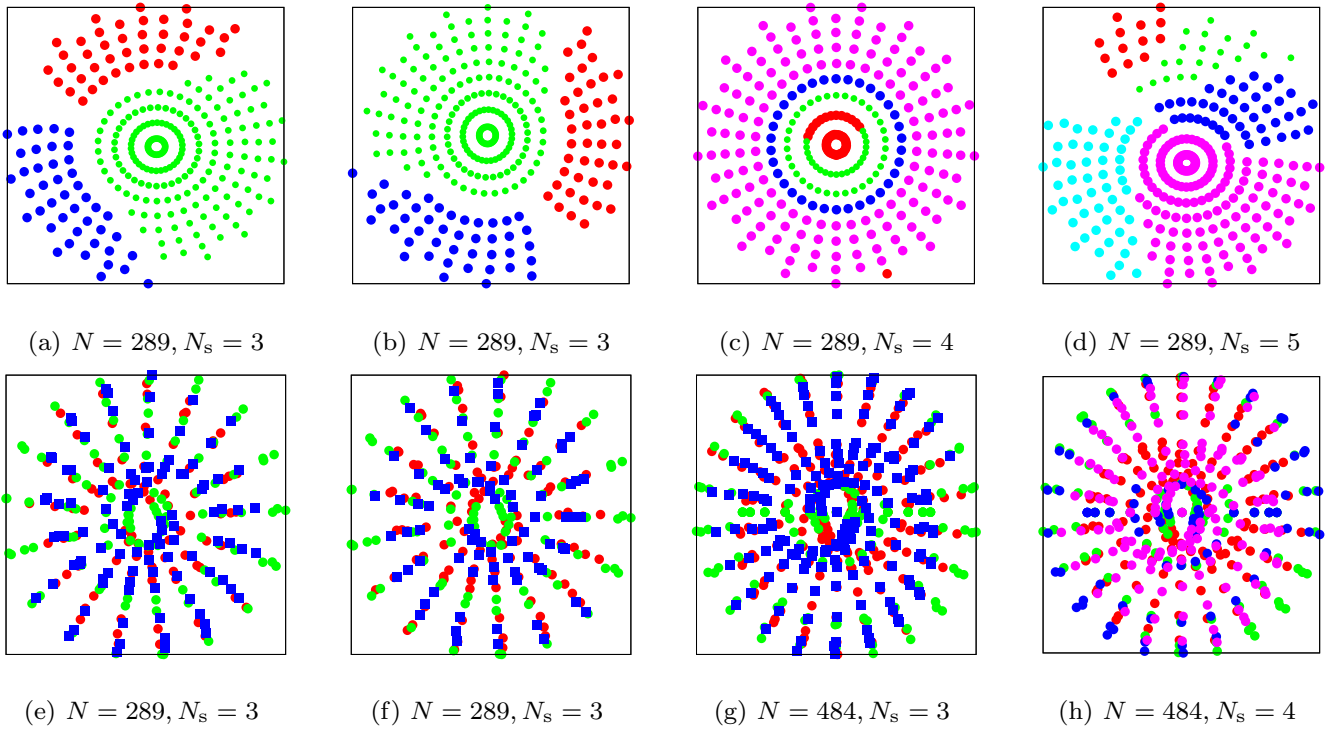}      
\caption{The variability or multi-fold degeneracy (in the physical sense) 
of the 2D projections. The alternative charts (each belonging to the specific unique categorical variable
$s_i$) of 3D half-sphere: see the parts (a), (b), (c), (d). The results obtained for 
the Klein bagel are plotted in the parts (e), (f), (g), (h). The projections are 
calculated for different initial conditions and stochastic realizations. The identical 
colors correspond to the categories: {\em red} ($s_i = 0$), {\em green} ($s_i = 1$), {\em blue} 
($s_i = 2$) or {\em pink} ($s_i = 4$). The mapping uncovered some key differences between 
the comprehensibility of the sphere and Klein bagel. The mapping and formation of 
the charts in the case of sphere is relatively well understandable. On the other hand, 
the interlinked and nested spiral structures corresponding to different categories 
represent the only projectable information about the Klein bagel.}
\label{Fig7}
\end{figure}

\begin{figure}
\centering
\includegraphics[width=1\linewidth]{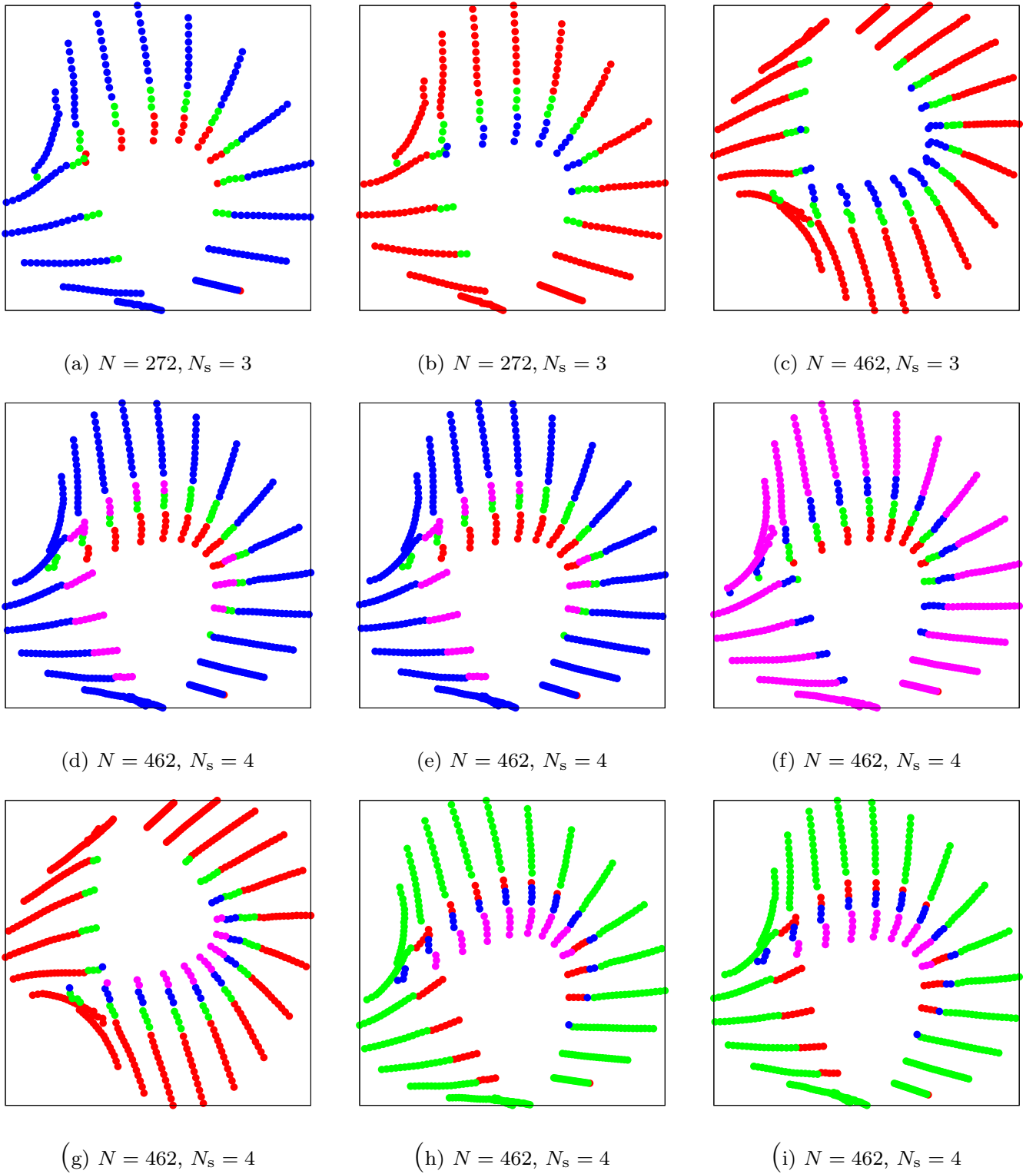}      
\caption{The alternative projections of the M\"oebius strip calculated for different initial conditions. 
Part (a): see the emergence of the middle strip which improves the orientation in the visualization of 
the structure. The denser data (c) better describe the higher variability, but they does not reveal some excessive differences. 
Note that difference between $s_i=0$ and $s_j=2$ domains (accompanied by the color difference) is only apparent 
since the locations of the classes $0, 2$ are 
replaceable because of their unique metric inter-class 
distance multiplicative factor $ \sim 1 + H |s_i-s_j| = 1 + 2 H $ 
(see Eq.(\ref{eq:Hell})).} 
\label{Fig8}
\end{figure}

The optimization of Klein bagel is presented in the Fig.\ref{Fig6}. Qualitatively, the
optimization scenario is similar to the 6D modulated spiral, as well as to other simulations 
that we performed. The optimized projections of the half-sphere, Klein bagel, and full sphere
samples are depicted in Figs.~\ref{Fig7},~\ref{Fig8} and \ref{Fig9}, respectively.
We see that  these projections exhibit different levels of segmentation and compactness.
Although the sample of Klein bagel resists partitioning into the compact regions
of different categories, its spiral motif becomes more clear and plainly visible
(see Fig.\ref{Fig2}) after the application of proposed method. 

In addition to studies of artificial datasets we propose the explanation 
for empirical observations and their similarities. We focus on the epidemiological 
data of Hodgkin lymphomas for United states in the 2009-2010. The mortality data~\cite{CDC2010} 
contain absolute death counts by age splitted into five race dimensions:  white~(1), black~(2), 
Asian/pacific islander~(3), American Indian/Alaska native~(4), Hispanic~(5) ($D=5$). 
The significance of the race of the patient for determination of the risk and 
efficiency of treatment has been discussed in~\cite{Zaki1993,LRF2014}. Before the application 
of MDS, the values on each particular dimension have been standardized
to have zero mean and unit variance. Comparison of the application of classical MDS with 
our approach is depicted in Fig.\ref{Fig10}.
Interestingly, both the 2D mappings show, that the observations can be embedded into
one-dimensional manifold, which confirms the salient role of to age-related disease
incidence. It also reflects the fact that time instants can be arranged in
a one-dimensional manifold. In addition to classical MDS, our method also identified specificity
of the categories belonging to the age bands 20-25, 30-40. It means, that the
adaptive metrics enables to detect even small decline in the disease occurrence.

All the above examples lead to questions regarding the role of the number of
categories, $N_{\rm s}$. Thus, it would be interesting to mention manifolds which evidently cannot be mapped onto the plane. 
In other words, the intrinsic dimension is too high. As a simple example demonstrating 
this property may serve the maps of vertices of 
6-dimensional hypercube, which are projectable onto 2D only on the expense of
very high stress values (when the categorization absents). 
On the other hand, when the categorization via $s_i$ is applied to the 
sample of $\mathbb{R}^{6}$, the layered or slice projection 
structures are generated (see Fig.\ref{Fig11}). In agreement with 
intuitive expectations, the minima of $V_{\rm tot}$ deepen 
with the increase of $N_{\rm s}$. 

The following conclusions can be 
drawn from the numerical examples:
\begin{description}
\item[(i)] the system dynamics exhibits qualitative universal features
which are independent from the investigated datasets; 

\item[(ii)] the "collective" coordinates $H(t), V_{\rm tot}(t)$ 
resemble the phase portraits of the forced double-well harmonic 
oscillators subject to strong noisy disturbances due to EO presence. 
Interestingly, the occurrence of two local minima 
(one deep and one more shallow) seems to be a generic feature common to wide
class of data. The exception is found in the uniform case $N_{\rm s}=1$, 
where the dependence on $H$ and $s_i$ simply vanishes.  

\item[(iii)] during the initial search phase $V_{\rm tot}(t)$ suddenly drops. 
The subsequent adjustment yields gradual refinement of the double-minima structure. 

\item[(iv)] the local, slow convergence with slow detailed search is typical for 
the last optimization stage (in agreement with the optimization model,
its assumptions and expectations, see the dynamics Eq.(\ref{eq:Hlearn})).

\item[(v)] in the most of investigated cases deeper minimum corresponds to
negative $H(t^{\ast})$.  It means, that our algorithm tends to interpret inter-category distances as smaller 
then $d^{(P=2)}_{i, j}$ (due to $1 + H | s_i - s_j |$ factor). 
This can be explained by the requirement to 
find sufficiently big area for the projections 
of the most of dense data inputs.

\end{description}

Interesting question arises whether oscillations of the parameters,
such as those which determine $H_{\rm per}(t)$, can be replaced
by the complex dynamical models. The promising candidate for the
alternative HO optimization part is the chaotic discrete Duffing
oscillator~\cite{Junge2004} which was used in our numerical experiments.
Our choice to use chaos has been motivated by the works 
which show increased optimization efficiency of the numerical sequences
generated by means of the chaotic maps when compared to the random
sequences~\cite{Li1997}. As an example we used the 
Duffing map $ x(t+1) = y(t)$,  $ y(t+1) = - 0.2
x(t) $  $ +  2.75 y(t) - [y(t)]^3$,  where $x(t+1)$ 
has been used to substitute the role of $ H_{\rm per}(t)$ ($y(t)$ is the auxiliary variable and the constants $-0.2$, $2.75$ 
were chosen to belong to the chaotic regime). Chaos is often 
present in the nonlinear systems, thus many another variants of chaotic dynamical systems can
be used to improve our MDS approach. Our preliminary simulations 
did not confirm increased efficiency in comparison to the presented harmonic stimulation.
To assess the relevance of the chaotic models for the 
complex optimization ML problem one needs further intensive numerical
research that goes far beyond the scope of the present work.
In any case, the application of the idea of self-organization in combination
with chaos phenomena may constitute very interesting 
way to develop further ML studies.

\section{Discussion} 

ML plays an important and growing role in exploratory data analysis and machine learning.
In the paper we introduced biologically and physically inspired flexible variant of standard
MDS, which, as we have shown, is the effective tool for simultaneous mapping and categorization. 
Our proposal uses the orchestration of three stochastic optimization heuristics. 
We demonstrated, that optimization trajectory produces the phase portrait 
involving two well-separated minima. This monitoring properly illustrates and, at the 
same time, justifies the necessity of the more comprehensive and advanced routines in the optimization. 

Current ML theories are handled in more or less linear framework with too small influence of non-linearity 
to exploit emergent characteristics. The main contribution of the present paper 
(in comparison with traditional ML theories) consists in combining reasonable 
solution of the specific problems with promising non traditional approximate 
heuristic methods used in the area of the complex systems, statistical physics, optimization science 
and artificial biology. Although not practical for the applications in which the projections and partitions must 
be found rapidly, the approach seems to be successful at more detailed analysis of the selected manifolds. 
We have shown, that the flexibility of MDS can be improved using the adaptive metrics 
containing categorical independent variables. Overall, surprising results clearly reveal
unexpected behavior and delineate the domains where this robust technique can bring
valuable results. Increased flexibility has to be paid with a larger number of categories.
Although the current form of the algorithm enables categorization of the datasets into predefined 
number of categories, further research is needed to tune the number to match
the intrinsic dimension and local structure of manifolds. The information criteria
(such as BIC) have to be included as well to improve the overall classification performance. 

Non-equilibrium dynamical systems often show complex adaptive behavior called emergent properties. 
In the most of the datasets that we studied the emergence of the compact domains of the categories
has been observed. It should be emphasized that processes, where structural changes 
of initial disordered configurations/projections yield self-organized structures, 
significantly differ from the traditional forms of the programming and learning which claim 
to produce similar effects by using some explicit and user predefined criteria. 

A lot of open questions remains to be studied in the proposed computational scheme. 
A general open question is whether we should generalize the method to include information 
about more generic distance functions and more nuanced interpretations and categorizations.
Our analysis does not rule out other modified forms of the $ 1 + H |s_i - s_j|$ prefactor 
of the metrics [see Eq.(\ref{eq:Hell})]. The model we used is flexible enough to admit 
straightforward extensions. For example, the prefactor 
$ 1 + H | s_i - s_j | $ $ +  H_{\mbox{\tiny symmetry\,\,\,viol.}} ( s_i+s_j )$ 
may be used which violate the original symmetry 
of $| s_i - s_j |$ (i.e. symmetry between an object $s_i$ and its mirror 
$N_{\rm s}-1-s_i$) by adding, 
e.g., the term $H_{\mbox{\tiny symmetry\,\,\,viol.}} ( s_i + s_j)$ 
that prevent from the occurrence of non-uniqueness and degeneracy.

Our analysis leads to interesting application of the hysteretic optimization method 
which has relevance in modeling and adjusting of systems with global impact parameters. 
We have shown that the examples can be found in ML situations where high intrinsic dimension of manifold 
favors the charting of data during the projection process. In particular, we believe that our numerical 
experimentation might be instructive in the construction of the models of the collective 
behavior including features such as emergence and organization, which are topics of much 
interest in the current research. We foresee further potential applications to variety of ML problems 
that are formulated as optimization tasks. 

\renewcommand{\abstractname}{Acknowledgements} The authors would like to gratefully acknowledge 
Project CELIM (316310) "Fostering Excellence in Multiscale Cell Imaging" 
funded by European Community Seventh Framework Program FP7 EU, and European X-Ray Laser Project XFEL.  

\section{Appendix - list of the synthetic data structures} 

The appendix describes the list of three parametrizations I, II, III of 3d manifolds, 
used to generate datasets suitable for 
MDS variant of ML numerical experiments. 

\vspace*{2mm}

\noindent I. \underline{Half - Sphere} ($D=3$ approximated by $N=17^2 = 289$ data items)

\vspace*{2mm} 

\begin{figure}
\centering
\includegraphics[width=1\linewidth]{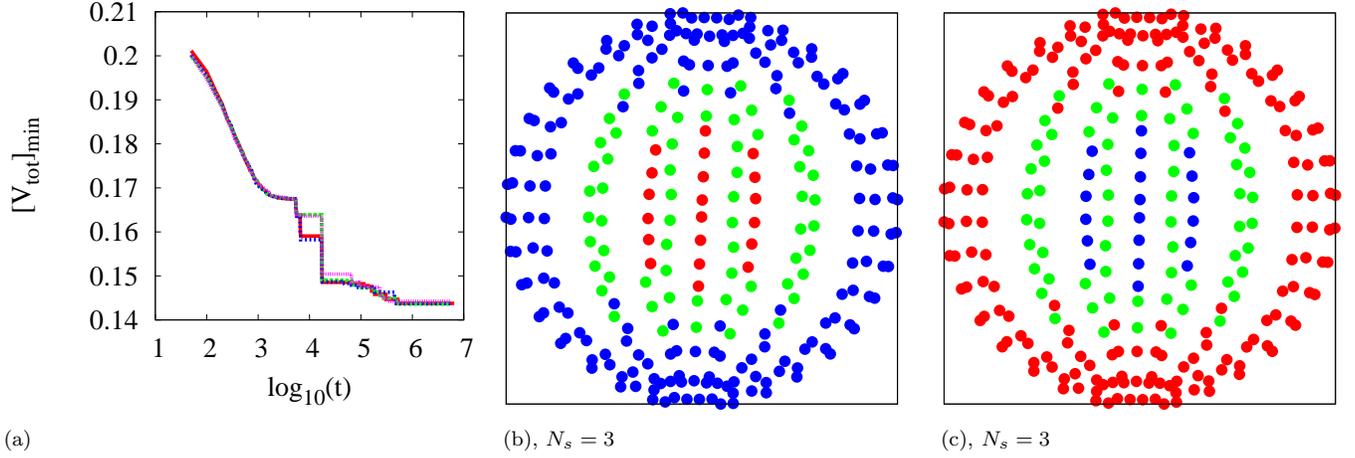}      
\caption{The demonstration of the two-fold degeneracy in the case of the categorization 
of the data drawn from the sphere. Our version of MDS is applied to data drawn from 
the 3D sphere. The part~(a) represent the actual minimum is recalculated for four 
independent simulation stochastic runs. Two examples - parts (b) and (c) of the configurations that exhibit the
invariance with respect to spin transformations:  $s=0$ (red)   $\rightarrow $ 
$s=2$  (blue) ; $s=1$ (green) $\rightarrow $ $s=1$  (green). The emergence
of distinguishing characteristics (see alternating vertical lines) of 
the front and back surface of sphere.}
\label{Fig9}
\end{figure}

\begin{figure}
\centering
\includegraphics[width=1\linewidth]{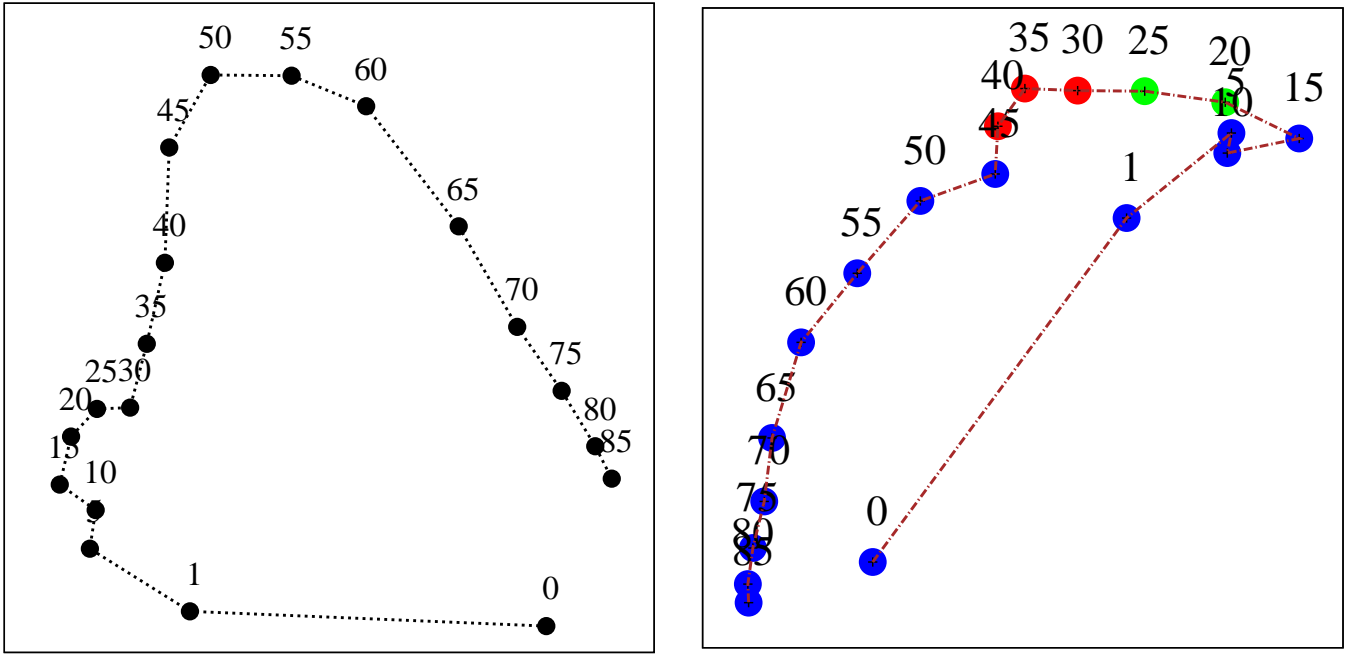}      
\caption{The MDS results obtained for 5-dimensional race-specific data 
for Hodgkin's lymphoma data. 
The comparison of classical metric MDS and our approach (which provides optimum 
$H=-0.064$, $V_{\rm tot}=0.026$. The effect of young - old age similarity closing 
the loop is remarkable. The key advantage of our methods is that it identifies the significant 
changes in the categories of the middle-aged adults.} 
\label{Fig10}
\end{figure}

\begin{figure}
\centering
\includegraphics[width=1\linewidth]{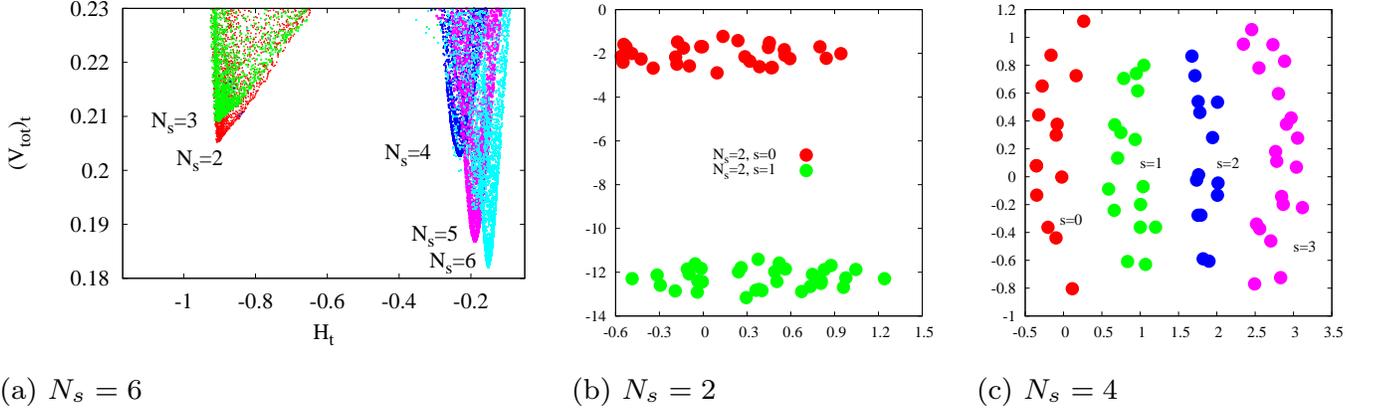}      
\caption{The influence of $N_{\rm s}$ integer parameter on the depth of the minimum
$V_{\rm tot}$ achieved (see part a) for dataset constructed 
from the vertices of 6-dimensional cube. The emergence of layered structures 
presented in the parts (b) and (c).}
\label{Fig11}
\end{figure}

The data are generated using 
\begin{eqnarray}
X_{i,1} &=&  \cos\phi_j\, \cos\theta_k\,, \qquad  X_{i,2} = 
\sin\phi_j\,\cos\theta_k\,, 
\\
X_{i,3} &=&  \sin\theta_k\,, 
\nonumber 
\\
\theta_k &=&  \pi (k/17)\,,\qquad \phi_j = 2 \pi (j / 17)\,\,.
\nonumber 
\end{eqnarray}
Here $\theta_k$ and $\phi_j$ denotes the sequences (samples) in azimuthal and polar coordinates, respectively. The index $i \equiv  i (j,k)$ represents 
the enumeration mark of $(j,k)$ elements of the Cartesian product
\begin{equation}
CP_{\rm I} \equiv   \{\,\, (j, k)\,;\, \, \,\,   j =  0, 1, \ldots, 16\,;  \,\,\, k =  0, 1, \ldots, 16\, \}\,.
\end{equation}
Thus for example: $i(0,0)=1$ (it means that here $j=0$, $k=0$), $i(0,1)=2$ (here $j=0$, $k=1$), 
$i(0,2)=3$, $\ldots$ $i(1,0)=18$, $i(1,1)=19$, $i(1,2)=20$, $\ldots$ $i(16,14)=287, i(16,15)=288, i(16,16)=289$. 
The analogous notation is used in the case of datasets II and III. Note that sample of the full 
sphere which is projected in Fig.(\ref{Fig9}) is created by replacements 
$\cos\theta_k\rightarrow \sin \theta_k$, $\sin\theta_k \rightarrow \cos \theta_k$. It means 
that data object we call "full sphere" includes the same number of data
inputs as the data object "half sphere". 

\vspace*{3mm}

\noindent II. \underline{M\"oebius strip} ($D=3$; $N=16 \times 17 = 272$ data items; we examined also denser data variant $N = 462=22 \times 21$); 

\vspace*{1mm}

\begin{eqnarray} 
X_{i,1} & = &  \left[ 1  + (v_j/2) \sin(u_k/2) \right]\,  \cos u_k\,, 
\\
X_{i,2} & = &  \left[ 1  + (v_j/2) \cos(u_k/2) \right]\,  \sin u_k\,,  
\nonumber 
\\  
X_{i,3} &=&   ( v_j /2 ) \sin(u_k/2)\,, 
\nonumber 
\\
v_j  &=&  1 - 2 (j / 17)\,,\qquad  u_k  =  2 (k / 17) \pi\,. 
\end{eqnarray} 
Again $i  \equiv  i(j,k)$ enumerates $CP_{\rm II} \subset CP_{\rm I}$,  
where 
$ CP_{\rm II}  \equiv  \{\,\,  (j, k)\, ; \, \, \, j =  0, 1,  
\ldots, 16\,;  \,\,\, k = 1, 2, \ldots, 16\,  \}\,$ 
slightly differs from $CP_{\rm I}$. 

\vspace*{4mm}

\noindent III. \underline{Klein bagel} ($D=3$; $N=289$ data items; denser dataset variant includes $N=484$ items;) 

\vspace*{2mm}

The manifold is homeomorphic to the well known Klein bottle. The data are generated using 
\begin{eqnarray}
X_{i,1}  &=&  \left[ a_{\rm R} + \cos( \theta_j/2)   \sin v_k - \sin(\theta_j/2) \sin(2 v_k)  \right] \cos\theta_j \,,  
\\
X_{i,2}  &=&  \left[ a_{\rm R} + \cos( \theta_j/2)   \sin v_k - \sin( \theta_j/2) \sin(2 v_k)  \right] \sin\theta_j \,,
\nonumber 
\\ 
X_{i,3}  &=&  \sin(\theta_j/2)  \sin v_k + \cos(\theta_j/2) \sin(2 v_k)\,, 
\nonumber 
\\
\theta_j &=&  2\pi  j/ 17\,, \qquad  v_k = 2 \pi k/17 \,. 
\end{eqnarray} 
The model depends on the single parameter 
we choose $a_{\rm R}=1.5$. Here $i \equiv i(j,k)$ enumerates 
set $(j,k)  \in CP_{\rm I}$. 


\newpage 

\end{document}